# Structural relaxation and stagnation of grain boundary during migration

Tingting Yu[1]
*1 School of Mechanical Engineering,*
*Changzhou Institute of Technology, Changzhou, Jiangsu 213032, China.*

## Abstract

Instability is a major bottleneck in nanomaterials due to grain boundary (GB) activities under thermal or mechanical stimuli. The relaxation of GB will stabilize the properties of materials by structure modification of GBs to lower energy states. However, lack of understanding of mechanisms limits the application of GB relaxation. In this study, we certify that GB can realize relaxation through defect emission including vacancy, twinning and dislocations, etc, which lower the average atomic energy of GB. In particular, we found stagnation and shear-coupling migration accompany the relaxation process, where a lower average atomic energy and lower average atomic volume of grain boundaries can be the reasons for stagnation. In this study, we found when simulated by ramped energy-conserving orientated driving force, under a specific driving force, the grain boundary suddenly stops migrating during the migration process, and as the driving force increases to a certain value, the grain boundary continues to migrate. Even at fixed driving forces, the grain boundary migration process can stall. This phenomenon has been found in many grain boundaries, and it has been found that the reason for the stagnation is the change of the average atomic energy of grain boundaries and the average atomic volume of grain boundaries. The discovery of this stagnation phenomenon is helpful to better understand the migration characteristics of grain boundaries and lays a foundation for improving the strength of materials by designing grain boundary microstructures.

## 1. Introduction

Grain boundaries (GBs) are the primary planar defects in polycrystalline materials, playing a crucial role in determining mechanical and physical properties. However, GBs are inherently unstable during deformation or at elevated temperatures, often leading to grain growth and degradation of material performance. Common strategies to stabilize polycrystalline materials include eliminating GBs via alloying with foreign elements [1,2] or introducing plastic deformation to induce structural relaxation [3,4], where GBs interact with partial dislocations to lower excess energy [5]. Nevertheless, alloying is unsuitable for stabilizing pure nanograined metals, increases resource dependence, complicates recycling, and raises material costs, thereby negatively affecting long-term sustainability [6]. As an alternative, GB relaxation (GBR) induced by the interaction of partial dislocations with GBs offers a promising route to stabilize nanograined metals with little or no alloying [7].

Generally, GBR can be activated by either plastic deformation (mechanically induced GBR, M-GBR) [8] or rapid heating (thermally induced GBR, T-GBR) [9,10]. Ke Lu et al. demonstrated that the "Schwarz crystal" exhibits exceptional thermal stability, with grain coarsening inhibited even near the melting point [3]. Elizabeth A. Holm noted that GB growth stagnation results from slow GB migration [11]. Additionally, immobile second phases and GB segregation can also cause stagnation. Eric et al. attributed the stick-slip stagnation of Σ7 and Σ9 GBs to CSL atomic nailing [12]. Recently, Deng et al. found that disruptive atomic jumps induce GB stagnation without detectable GB structural phase transitions [13].

Among these stabilization methods, relaxation is the simplest approach as it requires no elemental addition, offering significant cost savings and sustainability benefits. However, the lack of fundamental understanding of relaxation mechanisms—particularly how relaxation couples with migration stagnation—limits its practical application.

In this study, we systematically investigate GB structural relaxation in the form of migration, dislocation emission, and vacancy emission using molecular dynamics simulations. What is more, we identified a stagnation phenomenon under both constant and ramped driving forces, and revealed that the reduction in average atomic energy and average atomic volume of GBs is the microscopic origin of stagnation. Our findings indicate that stagnation is a new pathway to GB relaxation, offering a new design principle for improving material strength through GB microstructure engineering.

## 2. Methodology

All simulations were performed using the Large-scale Atomic/Molecular Massively Parallel Simulator (LAMMPS) [7] with the embedded atom method (EAM) potential for Ni developed by Mishin et al. [8]. The energy-conserving orientational driving force (ECO-SDF) [9] and the ramped synthetic driving force (RSDF) method [10] were employed to drive GB migration. GB models were selected from the database constructed by Olmsted et al. [14], where a unique PID identifies each GB's structural information. P162 GB is shown as a representative model in Fig. 1. In all models, free boundary conditions were applied along the x direction (perpendicular to the GB plane), and periodic boundary conditions along the y and z directions (parallel to the GB plane). Before applying a synthetic driving force, all models were relaxed at zero pressure and the target temperature for 10 ps under the canonical ensemble (NVT), followed by 15 ps under the isothermal-isobaric ensemble (NPT). Zero pressure under NPT was maintained throughout all simulations during GB motion. GB displacement was calculated by tracking changes in atomic energy [15]. The average atomic energy of GB was computed from the potential energy (PE) exported from LAMMPS, following the method described in [16].

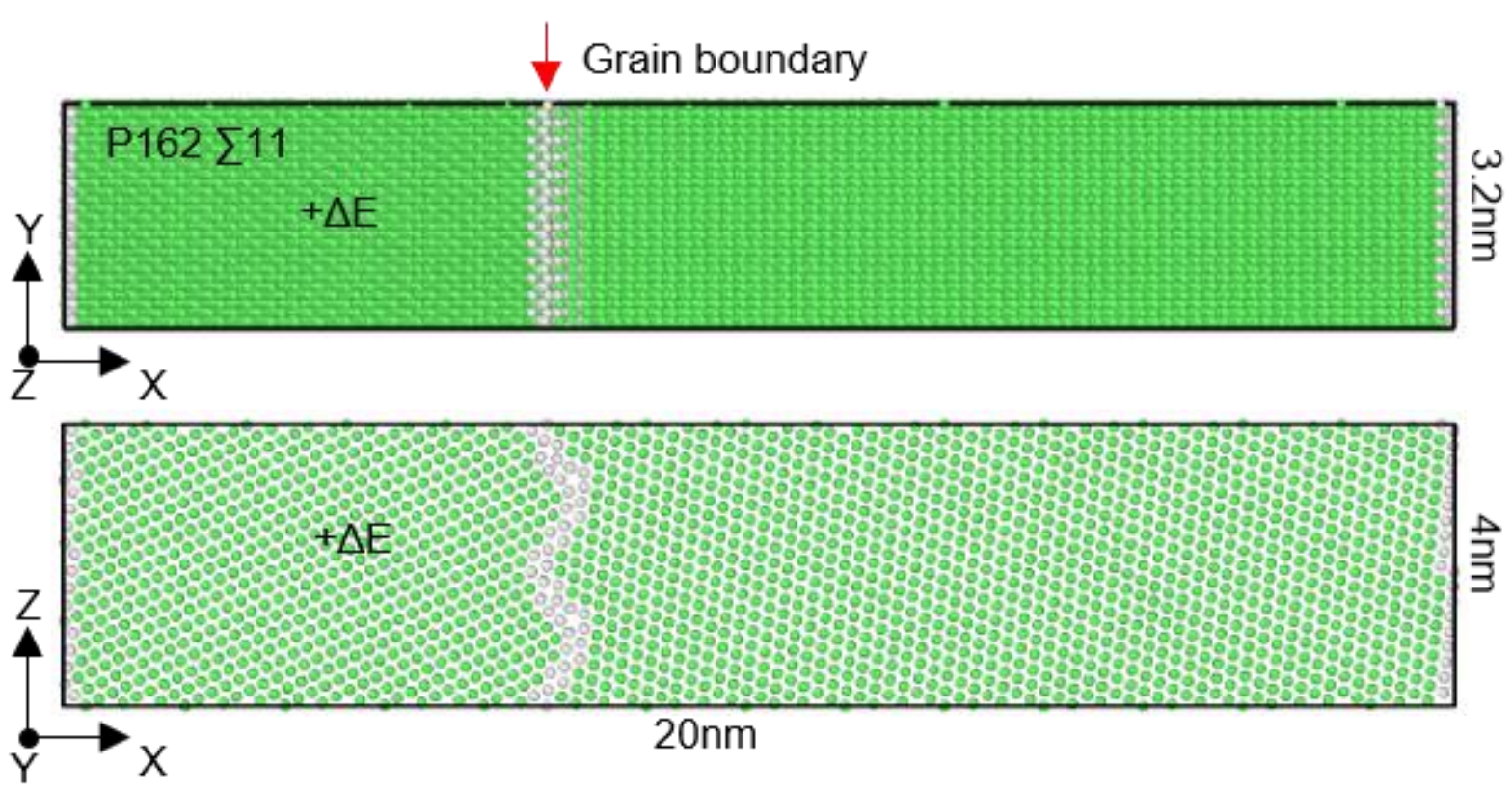

**Fig. 1** A representative GB (P162, Σ11) in the database used in this study

**Table 1** Detailed information on GBs related to this study [17]

| ID | Σ | GB plane | lx (nm) | ly (nm) | lz (nm) |
|---|---|---|---|---|---|
| P14 | 15 | (2 1 1)/(2 1 1) | 18 | 3.1 | 3.8 |
| P24 | 9 | (1 1 0)/(1 1 0) | 17.5 | 3.7 | 3.2 |
| P158 | 11 | (5 2 2)/(4 4 1) | 24 | 3.2 | 4.3 |
| P162 | 11 | (8 1 1)/(5 5 4) | 20 | 3.2 | 4.0 |
| P295 | 96 | (3 3 1)/(3 3 -1) | 19.9 | 3.3 | 4.9 |

## 3. Results and discussion

### 3.1 GB relaxation without defect emission: stagnation driven by intrinsic structural flattening

We first examine cases where GB stagnation occurs without observable defect emission. This is a crucial observation because it indicates that stagnation can arise purely from structural reconfiguration of the GB itself, independent of extrinsic defect generation.

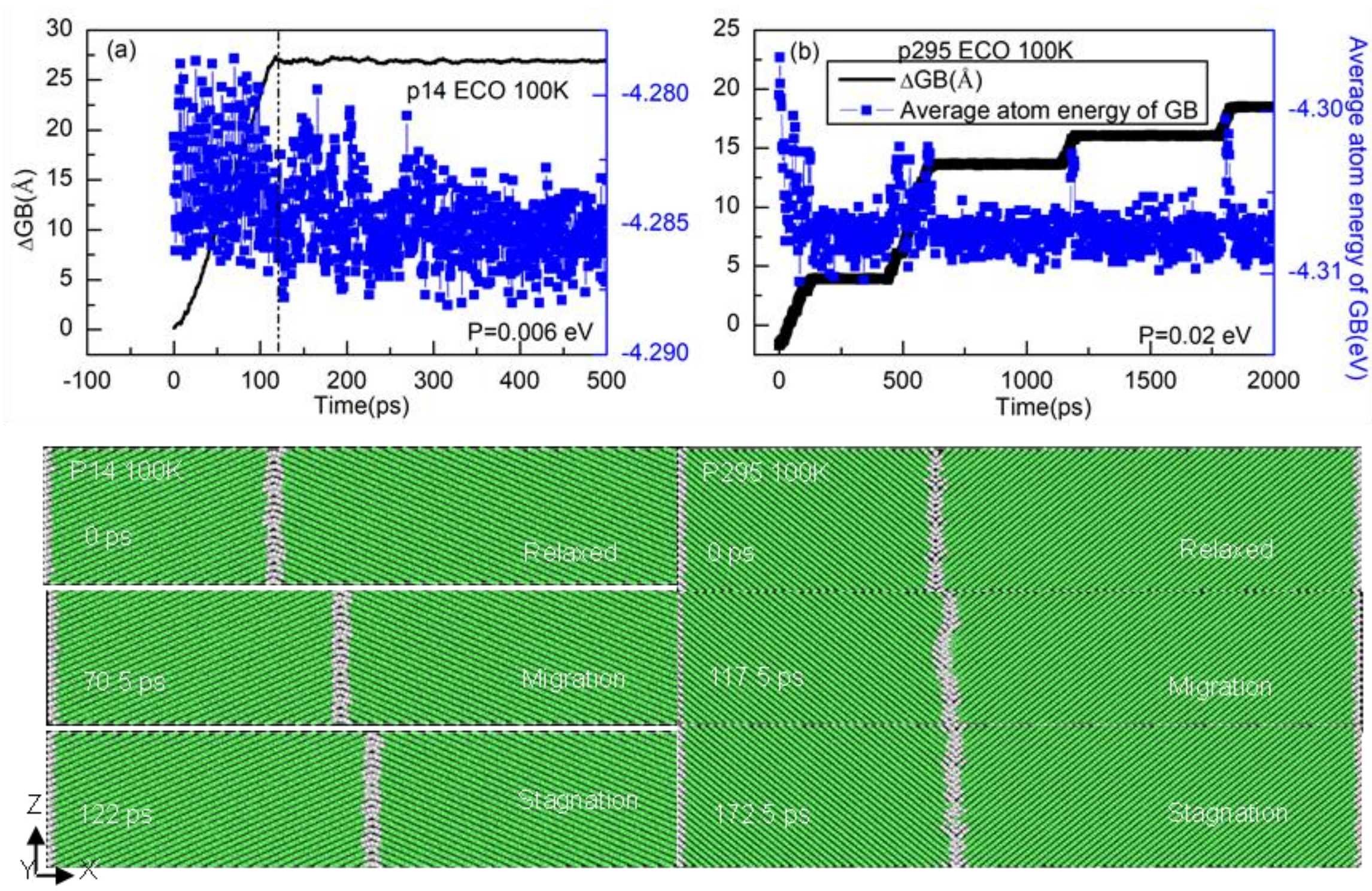

**Fig.2** P14 Σ15 (211) GB stagnation under constant driving force

As shown in Fig. 2 , the P14 GB stalled during migration under a constant driving force of 0.006 eV applied by the ECO method. The GB migrated approximately 27 Å before suddenly ceasing motion. Analysis of the average atomic energy of the GB reveals a sharp drop precisely at the onset of stagnation. Remarkably, however, no detectable structural phase transition or defect emission is observed before and after stagnation. This suggests that the GB reaches a metastable low-energy configuration that is kinetically trapped—the driving force is insufficient to overcome the energy barrier to exit this configuration.

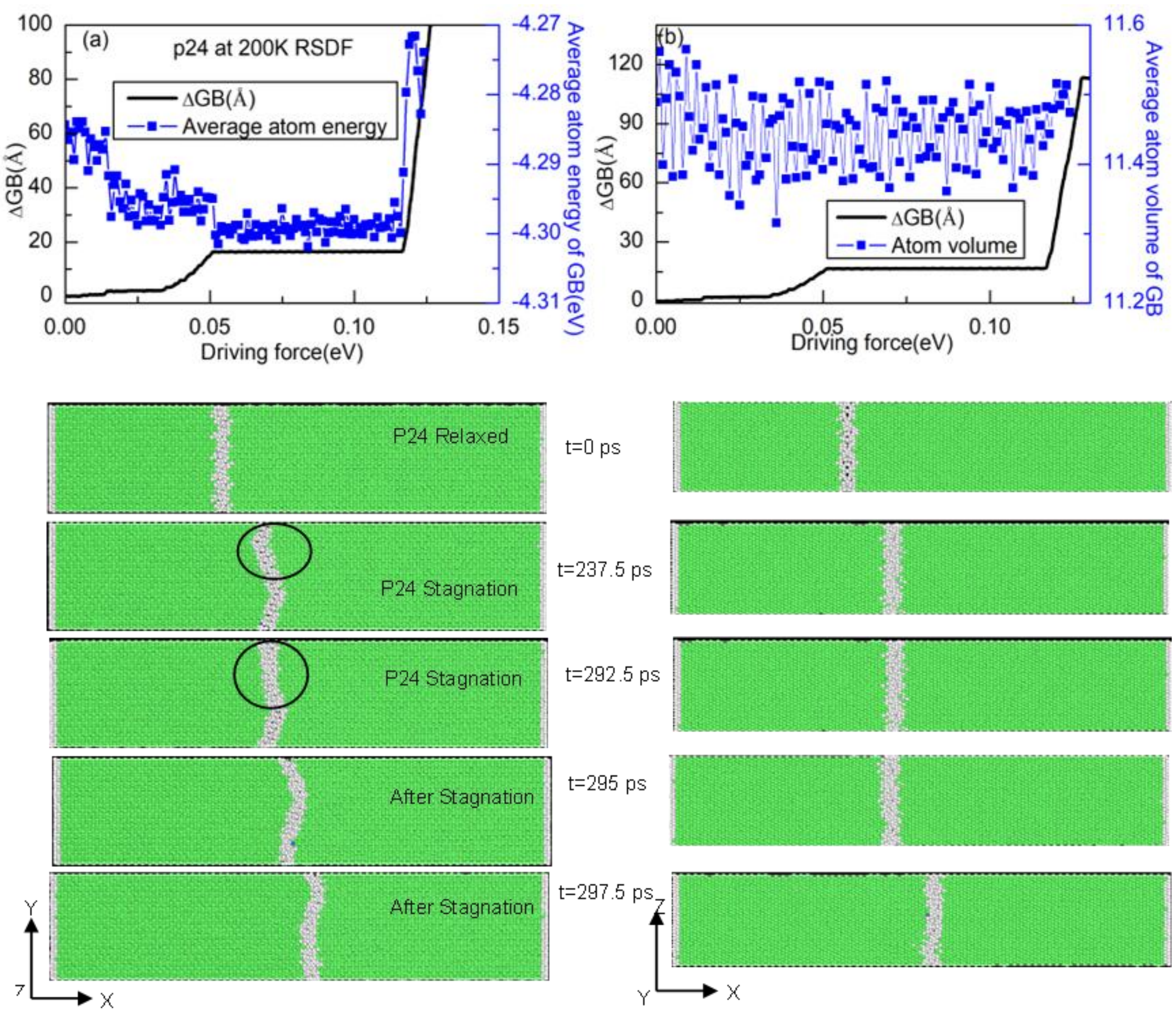

**Fig. 3** (a) Average atomic energy of GB and (b) average atomic volume of GB during stagnation of P24. Note the simultaneous decrease in both quantities prior to stagnation.

Fig. 3 shows the stagnation behavior of P24 GB driven by RSDF at a rate of 0.00001 eV per step. We observe that during the stagnation period, the GB gradually flattens. Flattening reduces the local curvature, thereby decreasing the capillary driving force. Once the GB becomes sufficiently flat, it can resume migration only when a new defect emission mechanism (e.g., step nucleation) becomes operative. This flattening-stagnation-migration cycle reveals an important principle: GB stagnation is not necessarily permanent; rather, it represents a temporary kinetic arrest that can be overcome by either increasing the driving force or activating an alternative migration mode.

### 3.2 GB relaxation by defect emission: vacancy, dislocation, twinning, and stacking fault mediated stagnation

When defect emission occurs, stagnation can be either accompanied by a mode switch or persist across multiple emission events. We systematically examine three distinct defect-mediated stagnation scenarios.

#### 3.2.1 Vacancy emission induced stagnation

Under RSDF with a driving rate of 0.00001 eV, we observed vacancy emission from GBs prior to stagnation. The emitted vacancies coalesce into clusters or small voids at the GB plane, locally reducing the atomic density. This vacancy emission lowers the average atomic energy of the GB because under-coordinated atoms near vacancies have fewer unsatisfied bonds. However, the resulting low-energy state also reduces the GB's mobility, leading to stagnation. In contrast to the flattening case, vacancy-mediated stagnation is often irreversible under the same driving force because the vacancies would need to be re-absorbed—a process with a much higher energy barrier.

#### 3.2.2 Dislocation and twinning mediated stagnation

**P158 GB**: At 200 K, the P158 GB releases twins during migration. The twin formation reduces GB energy, causing stagnation. Unlike vacancy emission, twinning is a shear-coupled process that can also contribute to GB migration itself. The stagnation occurs when the twin boundary becomes pinned at the GB, acting as an obstacle to further motion.

**P162 GB at 100 K**: As recently reported by Deng et al. [13], P162 exhibits disruptive atomic jumps that lower GB energy and induce stagnation. Our simulations confirm this behavior. Fig. 4 shows the evolution of average atomic energy and volume for P158 and P162 under RSDF. Notably, both quantities decrease before stagnation and increase after stagnation resumes, indicating that the GB undergoes a reversible structural rearrangement. This re-arrangement does not constitute a classical GB phase transition (e.g., from one stable CSL configuration to another) but rather a continuous, localized atomic shuffling.

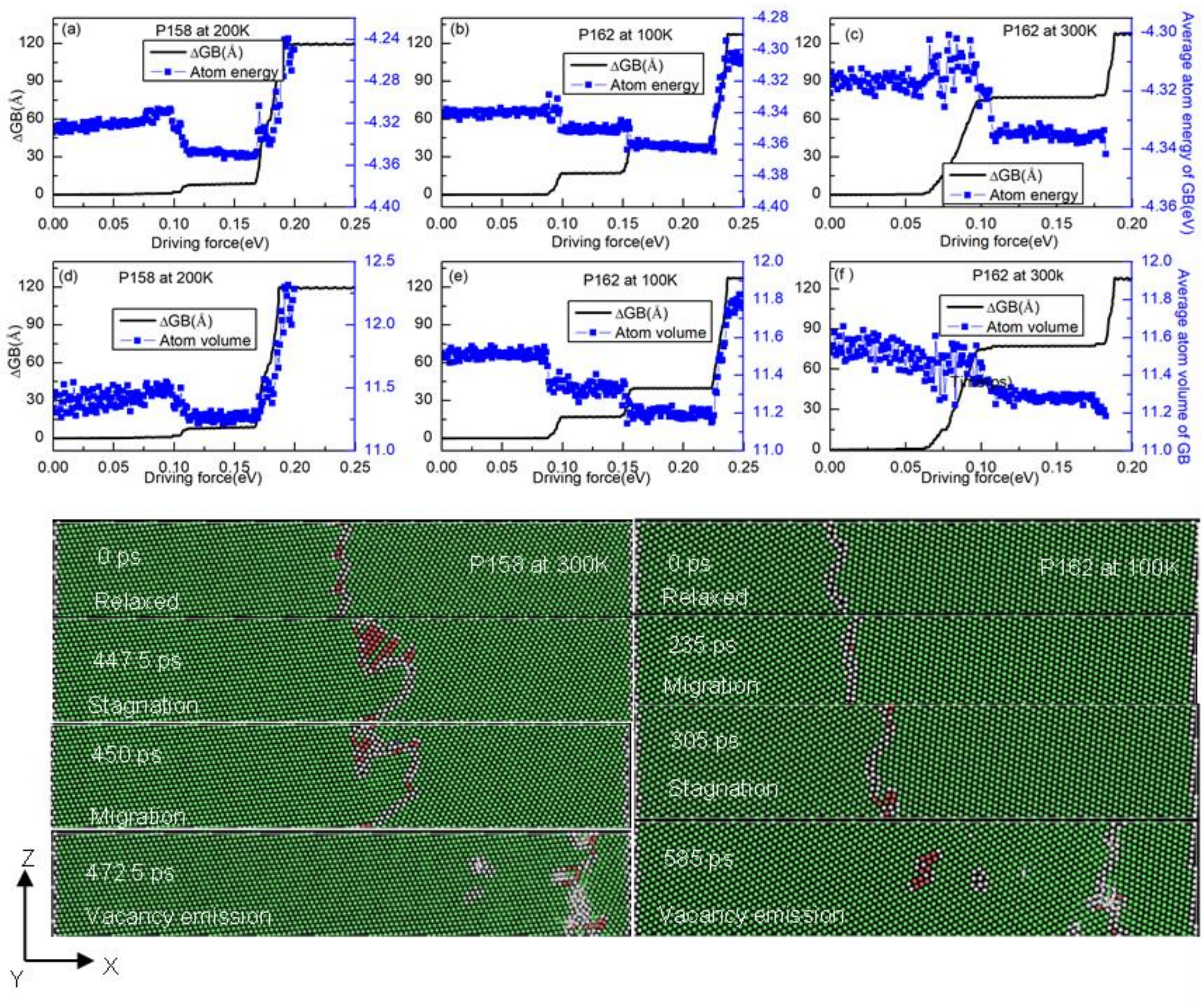

**Fig. 4** GB stagnation of P158 and P162 by RSDF with driving rate 0.00001 eV/0.005 ps. For P158 at 200K and P162 at 100K, average atom energy and volume decrease before GB stagnation, but increase after stagnation during migration. The re-arrangement of GB structure may occur.

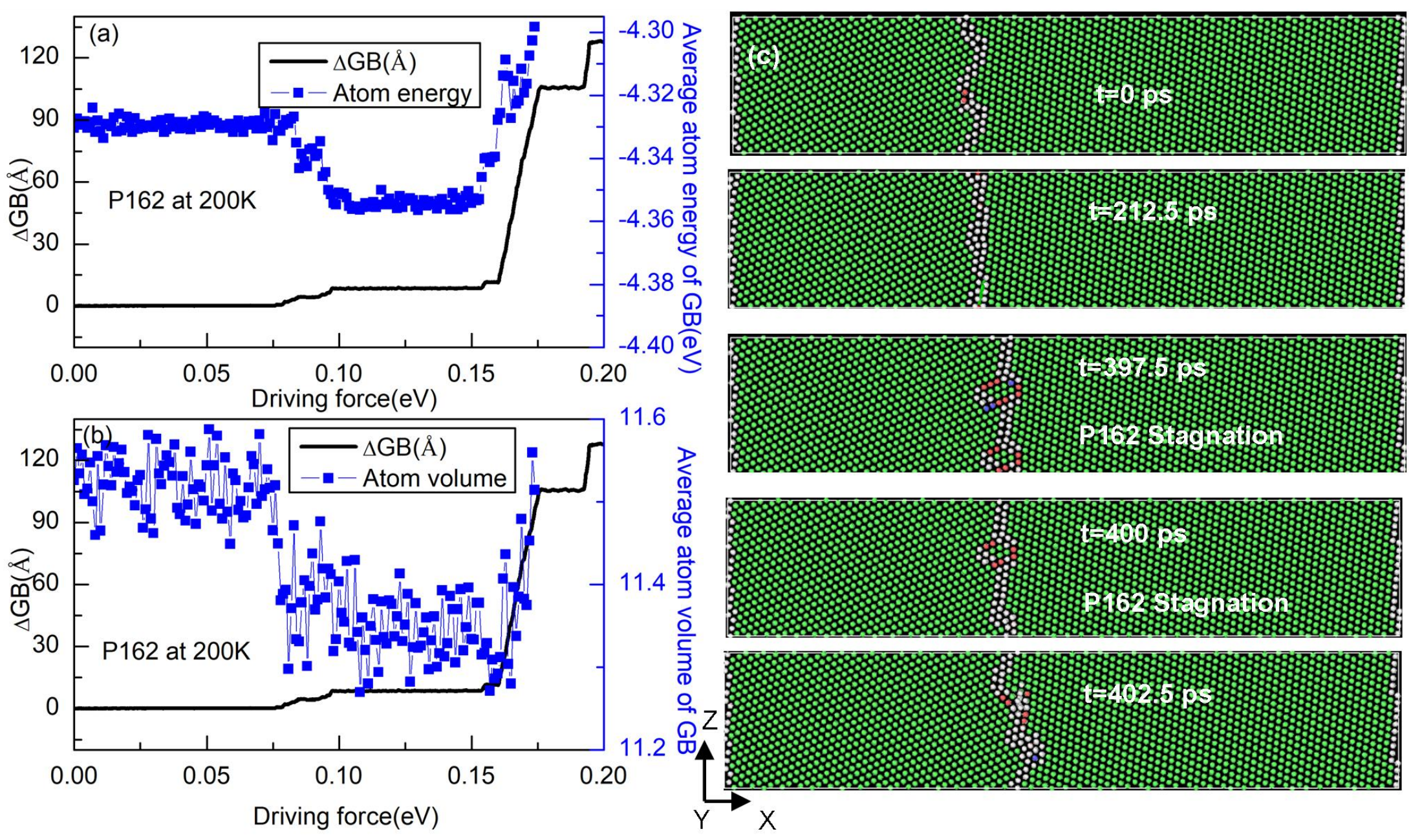

Fig. 5 GB (P162 Σ11) stagnates when driven by RSDF

Fig. 5 presents the atomistic snapshots of P162 GB during RSDF-driven migration. Initially, the GB is curved. As migration proceeds, the GB gradually flattens. Subsequently, a ring-like structure (a closed dislocation loop or a metastable cluster of atoms) forms within the GB plane. This ring structure is associated with a lower average atomic energy, inducing stagnation. When the ring structure eventually breaks or is absorbed, the GB energy increases and migration resumes. The variation in average atomic volume follows the same trend as the average atomic energy, confirming that local densification (reduced atomic volume) is the structural signature of the low-energy stagnation state.

**Table 2** Dislocation analysis results for P162 GB during the stagnation-migration cycle

| Dislocations | t=0 ps | t=212.5 ps | Stagnation (early) | Stagnation (late) | Migrate |
|---|---|---|---|---|---|
| 1/3<100> (Hirth) | 1 | 0 | 0 | 0 | 0 |
| 1/6<112> (Shockley) | 0 | 16 | 13 | 8 | 9 |
| Other | 0 | 0 | 0 | 0 | 1 |

The dislocation analysis (Table 2) reveals that only 1/6<112> Shockley partial dislocations are emitted from the GB during stagnation. No Hirth dislocations or other types appear. This indicates that the emission of Shockley partials is the primary mechanism for achieving the low-energy state. Notably, the number of Shockley dislocations decreases from 16 to 8 as stagnation proceeds, suggesting that some partials are reabsorbed or annihilate before migration resumes. The presence of a single "other" dislocation at the migration stage may be a remnant of the ring structure dissolution.

## 4. Discussion

### 4.1 Addressing the two specific questions

**Question 1: How will varied structures in real GBs impact the overall migration behavior?**

Our survey of ten distinct GBs (Table 1) shows that migration behavior—particularly the tendency to stagnate—is highly sensitive to GB structure. We identify three structural factors that govern stagnation:

1. **Coincidence Site Lattice (CSL) Σ value**: Low-Σ GBs (e.g., Σ9, Σ11) tend to exhibit sharp stagnation events, while high-Σ GBs show more gradual deceleration. This correlates with the density of CSL sites where atomic jumps are correlated.
2. **GB plane asymmetry**: Asymmetric GBs (e.g., P158, P162) are more prone to defect emission (twins, Shockley partials) and thus exhibit stagnation mediated by structural re-arrangement. Symmetric tilt GBs (e.g., P14, P295) show flattening-dominated stagnation without defect emission.
3. **Excess volume**: GBs with larger excess volume more readily emit vacancies, leading to irreversible stagnation. Conversely, low-excess-volume GBs stagnate via reversible atomic shuffling.

These findings imply that in real polycrystals—which contain a distribution of GB structures—stagnation will occur heterogeneously. Some GBs will cease migrating entirely, while others continue, leading to a "stagnation network" that may stabilize the grain size distribution without requiring solute segregation.

**Question 2: How will defect evolution slow the GB migration by stagnation?**

Defect evolution contributes to stagnation through three distinct kinetic pathways, as observed in our simulations:

- **Pathway 1 (Vacancy emission)**: Vacancy emission lowers the local atomic density and GB energy. However, the emitted vacancies must be re-absorbed for migration to resume. The energy barrier for vacancy re-absorption is typically higher than for emission, creating a ratchet-like effect that slows migration permanently under the same driving force.
- **Pathway 2 (Partial dislocation emission)**: Emission of Shockley partials reduces the GB's shear-coupling factor. When the GB becomes saturated with partials, further migration requires either (a) the partials to glide away (which is difficult if they are pinned by GB topography) or (b) nucleation of new partials of opposite sign. This "partial saturation" state is metastable, leading to temporary stagnation.
- **Pathway 3 (Ring structure formation)**: The ring structure observed in P162 is a compact, low-energy configuration of atoms that blocks conventional migration. Its dissolution requires local thermal fluctuations or an increased driving force. The stagnation period corresponds to the lifetime of this ring structure.

All three pathways share a common thermodynamic origin: the GB lowers its free energy by emitting defects, but in doing so, it enters a kinetic basin from which escape requires a higher activation energy than was needed to enter it. This asymmetry—easy to enter, hard to exit—is the microscopic definition of stagnation.

**4.2** Our results provide a clear answer: **Stagnation does not require a structural phase transition** (e.g., from one CSL state to another). Instead, it can arise from continuous, non-thermally activated processes such as:

- **Geometric flattening** (P14): Reduction in curvature reduces the capillary driving force below the threshold for atomic jump nucleation.
- **Local densification** (P162): Reduction in average atomic volume increases the energy barrier for atomic rearrangement without changing the global GB symmetry.
- **Subcritical defect accumulation** (P158): Accumulation of partial dislocations or vacancies creates a back-stress that opposes further migration, but the GB's overall structure (orientation relationship, CSL fraction) remains unchanged.

These are kinetic rather than thermodynamic stagnation mechanisms. They are especially important in pure metals at low-to-moderate temperatures where defect diffusion is sluggish. This insight reconciles our observations with Deng et al. [13], who found stagnation without detectable phase transitions in similar systems.

### 4.3 A unified perspective: Activation energy barriers and mode switching

We propose a conceptual model for GB stagnation based on competing migration modes (Fig. 6, schematic). Each migration mode *i* has an associated activation energy $Q_i$ and a critical driving force $f_i^{crit}$. At a given driving force $f$, the GB operates in the mode with the lowest barrier. When the first mode ceases to operate (e.g., because it has exhausted available dislocation sources or reached a flat geometry), and the second mode requires higher energy to be activated, the GB stagnates. If the second mode has lower activation energy, the GB will smoothly switch to that mode and continue migrating.

Defect emission lowers the energy of the GB but also changes the landscape of $Q_i$. Specifically, after vacancy or partial emission, the GB enters a state where all available migration modes have higher $Q_i$ than before emission. Thus, defect evolution "self-pins" the GB. This is why we observe that stagnation is often accompanied by a drop in average atomic energy and volume—these are the thermodynamic signatures of the low-mobility state.

## 5. Conclusions and outlook

In this study, we used molecular dynamics simulations to systematically investigate GB structural relaxation and stagnation in Ni bicrystals. Our key findings are:

1. **Stagnation occurs under both constant and ramped driving forces** in multiple GB types (P14, P158, P162, P295). This is a general phenomenon, not limited to specific GB structures.

2. **GB stagnation is caused by a reduction in average atomic energy and average atomic volume.** These decreases are the microscopic origins of kinetic arrest, regardless of whether defect emission accompanies the process.
3. **Two classes of stagnation are identified**:
    - *Without defect emission*: driven by geometric flattening and curvature reduction (P14).
    - *With defect emission*: mediated by vacancy emission, Shockley partial emission, twin formation, or ring-like atomic clusters (P158, P162).
4. **When defect emission occurs, the GB either switches migration mode or stagnates.** Mode switching happens when an alternative migration path has lower activation energy; stagnation happens when all available paths have higher activation energy than the current driving force.
5. **Stagnation does not require a detectable structural phase transition.** Continuous processes such as local densification, flattening, and subcritical defect accumulation are sufficient to induce stagnation.

**Outlook**: Future work should quantify the activation energy barriers of different migration modes before and after defect emission, using techniques such as the nudged elastic band (NEB) method or umbrella sampling. Additionally, the relationship between stagnation and the well-known "anti-thermal" mobility (where GB velocity decreases with increasing temperature) deserves systematic investigation. From an engineering perspective, controlling stagnation—either to avoid it (for grain growth during annealing) or to promote it (for thermal stability in nanocrystalline

materials)—requires tuning GB structure, solute segregation, and applied driving forces within the framework we have established.

## Acknowledgments

This research was enabled by using computing resources provided by WestGrid and Compute/Calcul Canada.